\documentclass[
    ,final            
  ]
  {aipproc}
                                                                                
\layoutstyle{8x11double}

\def\D0{D\O}

\begin{document}

\title{Heavy Flavor Production at the Tevatron}

\author{Chunhui Chen 
\\ {\normalsize\sl (Representing the CDF and \D0 Collaboration)}
}{
address={Department of Physics and Astronomy, University of Pennsylvania, 
         Philadelphia, PA 19104, U.S.A.}
}

\begin{abstract}
Using a subset of the current Run\,II data, the CDF and 
\D0 have performed several measurements on heavy flavor 
production. In this paper, we present a new measurement 
of prompt charm meson production by CDF. We also 
report the latest CDF\,II measurements of inclusive $J/\Psi$ 
production and $b$-production without requirement of 
minimum transverse momentum on $J/\Psi$ and $b$-quark.
They are the first measurements of the total inclusive $J/\Psi$
and $b$ quark cross section in the central rapidity
region at a hadron collider. The results of $J/\Psi$ 
cross section as a function of rapidity, and 
$b$-jet production cross section measured by \D0
are also reviewed.

\end{abstract}

\maketitle


\section{Introduction}

Measurements of the production cross section of heavy 
flavor quarks ($c$ and $b$ quarks) in the $p\bar{p}$ collisions
provide us an opportunity to test the prediction based on 
the Quantum Chromodynamics (QCD). No only is QCD one of the
four fundamental forces of the nature, but also many searches 
of new physics beyond Standard Model require a good understanding
of the QCD background. 

The Bottom quark production cross section has been measured by both
the CDF and \D0 experiments~\cite{Abbott:1999se,Acosta:2001rz}
at the Fermilab Tevatron in $p\bar{p}$
collision at $\sqrt{s}=1.8\,\mbox{TeV}$, and found to be about three
times larger than the next-to-leading order (NLO) QCD 
calculation~\cite{Nason:1989zy,Albajar:1987iu}.
Since then, several theoretical explanation has been proposed to solve
the disagreement, such as the large contributions from the NNLO QCD
processes, possible contribution 
from ``new physics''~\cite{Berger:2000mp} etc. 
Recently, a more accurate description of $b$ quark fragmentation
has reduced the discrepancy to a factor 
1.7~\cite{Cacciari:2002pa}. Nevertheless, further
experimental measurements on the heavy flavor production are 
essential to shed light on the remaining clouds of this long
standing issue.

Since the end of Run\,I (1996), the Tevatron had underdone major
upgrade. The center of mass energy in the $p\bar{p}$ collision
was increased from 
$\sqrt{s}=1.8\,\mbox{TeV}$ to $\sqrt{s}=1.96\,\mbox{TeV}$.
The large luminosity enhancement dramatically increases the
discovery reach and moves the experimental program into 
a regime of precision hadron collider physics.
At the same time, both CDF~\cite{Blair:1996kx} and
D0~\cite{Abachi:1996hs} experimental collaborations 
significantly upgraded their detectors 
to enrich the physics capabilities, especially for the heavy flavor
physics program. The Tevatron Run\,II data taking period 
started in April 
2002, so far the accelerator has delivered about $300\,\mbox{pb}^{-1}$
integrated luminosities, both experiments collected more than 
$200\,\mbox{pb}^{-1}$ physics quality data. In this paper, we
will give a brief summary of the latest results
on heavy flavor production from Tevatron Run\,II.

\section{Prompt Charm Meson Production Cross Section}
In order to help understanding the discrepancy between 
measured Run\,I Bottom quark cross section measurement
and theoretical calculation, one can repeat
previous measurement with better control of experimental
uncertainties. An alternative approach is to examine
the production rate of another heavy quark.
Charm meson production cross~sections have not been measured
in $p\bar{p}$ collisions and may help to solve this 
disagreement.
For CDF\,II upgrade, one of the crucial 
improvement with respect to the CDF\,I is the implementation
of a Silicon Vertex Trigger (SVT)~\cite{Ashmanskas:1999ze}, 
which allows us to trigger on 
displaced tracks decayed from long-lived charm and bottom hadrons.
With this trigger, large amount of fully reconstructed 
charm meson has been collected, opening a new window for
heavy flavor production studies at hadron collider.
\begin{figure}
\scalebox{0.9}{\includegraphics{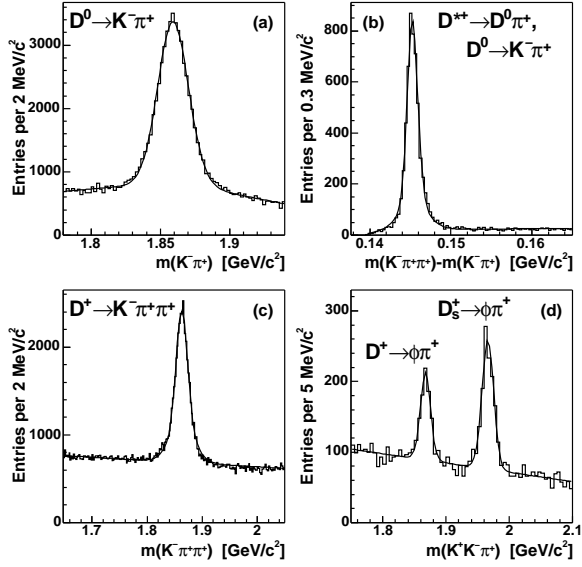}}
\caption{\label{fig:signal}
Charm signals summed over all $p_T$ bins:
(a) invariant mass distribution of  $D^0\to K^-\pi^+$ candidates;
(b) mass difference distribution of $D^{*+}\to D^0 \pi^+$ candidates;
(c) invariant mass distribution of $D^+\to K^-\pi^+\pi^+$ candidates;
(d) invariant mass distribution of $D^+\to\phi\pi^+$ and $D_s^+\to\phi\pi^+$
candidates.}
\end{figure}

Using just $5.8\pm0.3\,\mbox{pb}^{-1}$ data, CDF performs a
measurement of prompt charm meson production cross 
section~\cite{Acosta:2003ax}. 
The charm mesons are reconstructed in the following decay modes:
\mbox{$D^0\to K^-\pi^+$},
\mbox{$D^{*+}\to D^0\pi^+$} with \mbox{$D^0\to K^-\pi^+$},
\mbox{$D^+\to K^-\pi^+\pi^+$},
\mbox{$D_s^+\to \phi\pi^+$} with \mbox{$\phi \to K^+K^-$},
and their charge conjugate, as shown in Fig.~\ref{fig:signal}. 
We separate charm directly
produced in $p\bar{p}$ interaction (prompt charm) from charm
decayed from $B$ mesons (secondary charm)
using the impact parameter of the charm
candidate with respect to the primary interaction point. 
Prompt charm meson
points to the beamline and its impact parameter is zero.
The secondary charm meson candidate
does not necessarily point back to the primary vertex, it has 
a rather wide impact parameter distribution. The shape of 
the impact parameter distribution of secondary charm is obtained
from a Monte Carlo simulation. The detector impact parameter 
resolution is modeled from a sample of $K^0_S\to\pi^+\pi^-$ decays 
that satisfy the same trigger requirement. The prompt charm fraction is
measured as a function of charm meson $p_T$. Average over all
$p_T$ bins, $(86.6\pm0.4)\%$ of the $D^0$ mesons,
$(88.1\pm1.1)\%$ of $D^{*+}$,
$(89.1\pm0.4)\%$ of $D^+$, and
$(77.3\pm3.8)\%$ of $D_s^+$ are promptly produced 
(statistical uncertainties only). 

\begin{figure}
\scalebox{0.9}{\includegraphics{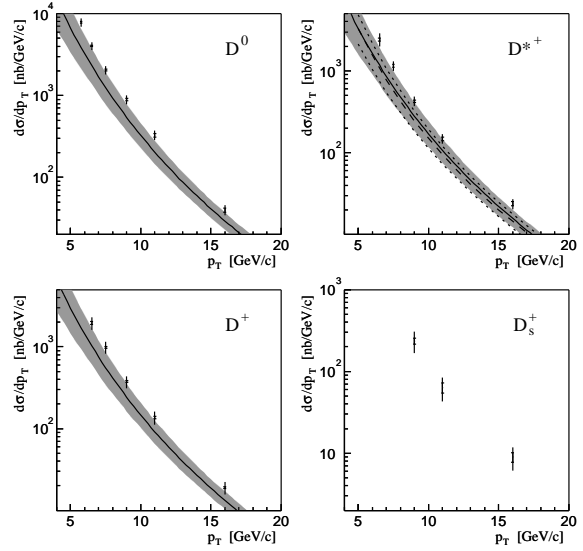}}
\label{fig:Dx}
\caption{
The measured differential cross section measurements for $|y|\leq1$, shown by the points.
The inner bars represent the statistical uncertainties; the outer bars are the
quadratic sums of the statistical and systematic uncertainties.
The solid curves are the theoretical predictions from Cacciari and Nason~\cite{Cacciari:2003zu},
with the uncertainties indicated by the shaded bands.
The dashed curve shown with the $D^{*+}$ cross section is the theoretical prediction
from Kniehl~\cite{kniehl};
the dotted lines indicate the uncertainty.
No prediction is available yet for $D_s^+$ production.
}
\end{figure}

\begin{figure}
\scalebox{0.9}{\includegraphics{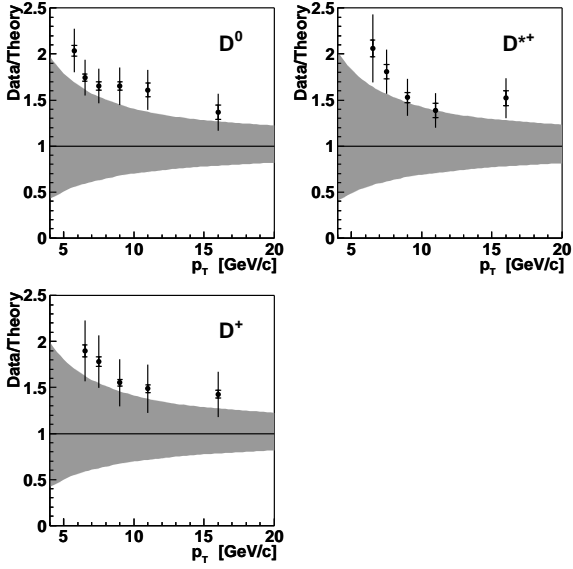}}
\caption{
Ratio of the measured cross sections to the theoretical calculation
from Cacciari and Nason.
The inner error bar represents the statistical uncertainty,
the outer error bar the quadratic sum of the statistical and 
systematic uncertainty. The hatched band represents the uncertainty 
from varying the renormalization and factorization scale.}
\label{fig:Dx_ratio}
\end{figure}

The measured prompt charm meson integrated cross section are found to be
$\sigma(D^0, p_T\geq 5.5\,\mbox{GeV}/c, |y|\le1)=13.3\pm0.2\pm 1.5\,\mu$b,
$\sigma(D^{*+}, p_T\geq 6.0\,\mbox{GeV}/c, |y|\le1)=5.2\pm0.1\pm 0.8\,\mu$b,
$\sigma(D^+, p_T\geq 6.0\,\mbox{GeV}/c, |y|\le1)=4.3\pm0.1\pm 0.7\,\mu$b and
$\sigma(D_s^+, p_T\geq 8.0\,\mbox{GeV}/c, |y|\le1)=0.75\pm0.05\pm 0.22\,\mu$b,
where the first uncertainty is statistical and the second systematic.
The measured differential cross sections are 
compared to two recent calculations~\cite{Cacciari:2003zu,kniehl},
as shown in Fig.~\ref{fig:Dx} and Fig.~\ref{fig:Dx_ratio}. 
They are higher
than the theoretical predictions by about 100\% at low $p_T$ 
and 50\% at high $p_T$. However, they are compatible within uncertainties.
The same models also underestimate $B$ meson production
at $\sqrt s=1.8\,$TeV by similar 
factors~\cite{Acosta:2001rz,Cacciari:2002pa,Binnewies:1998vm}.

\section{Inclusive $J/\Psi$ Production Cross Section}
\begin{figure}
\scalebox{0.4}{\includegraphics{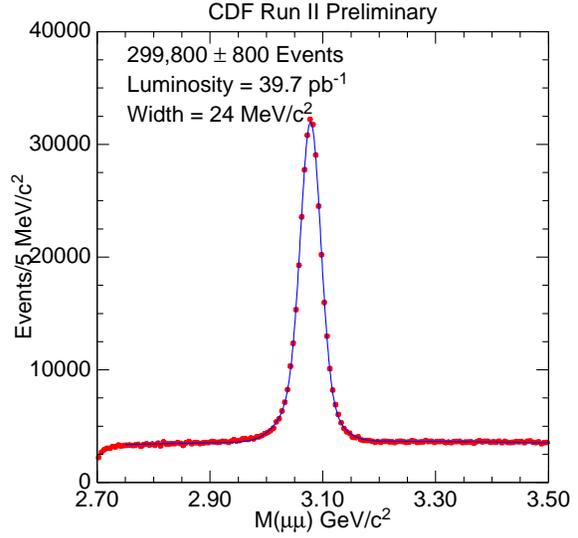}}
\caption{\label{fig:signal_Jpsi}{The invariant mass
distribution of triggered $J/\Psi$ events reconstructed 
in the $39.7\,\mbox{pb}^{-1}$ of CDF\,II data.}}
\end{figure}

\begin{figure}
\scalebox{0.4}{\includegraphics{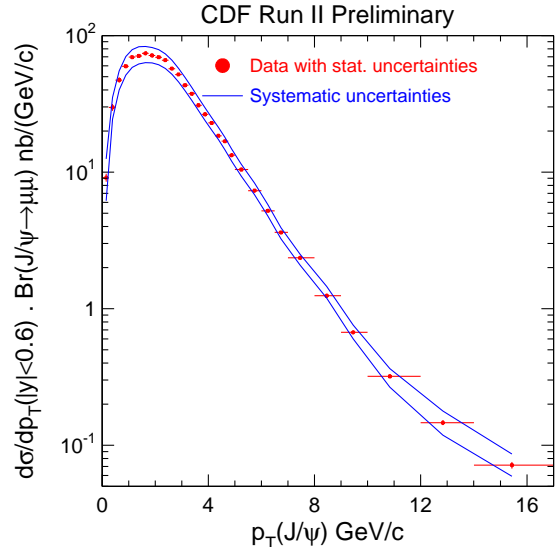}}
\caption{\label{fig:xsec_Jpsi}{The differential 
cross section for inclusive $J/\Psi$ as a function of $p_T$ 
with $|y|\le 0.6$.}}
\end{figure}

\begin{figure}
\scalebox{0.30}{\rotatebox{-90}{\includegraphics{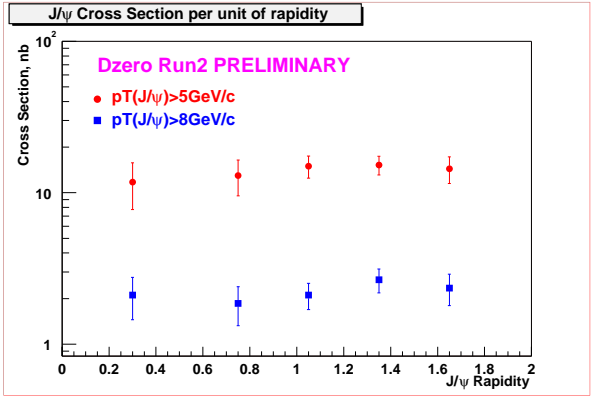}}}
\label{fig:xsec_Jpsi_D0}
\caption{The differential 
cross section for $J/\Psi$ as a function of $|y|$.}
\end{figure}

The CDF\,II detector has improved the dimuon trigger 
system with a lower threshold of $1.5\,\mbox{GeV}/c$.
This allows us to trigger the $J/\Psi\to\mu^-\mu^+$ event with 
transverse momentum all the way down to 
$p_T(J/\Psi)=0\,\mbox{GeV}/c$.
Using about $39.7\,\mbox{pb}^{-1}$ data, CDF reconstructed
 $299800\pm800$ $J/\Psi$ 
candidates (statistical uncertainty only), as shown 
in Fig.~\ref{fig:signal_Jpsi}.
A new measurement of the total inclusive $J/\Psi$ cross section 
in the central rapidity region $|y|\le0.6$ has been carried
out. This is the first measurement of the total inclusive
$J/\Psi$ cross section in the central rapidity region at a 
hadron collider. The differential
cross section is shown in Fig.~\ref{fig:xsec_Jpsi}, and the
integrated cross section is measured to be:
\begin{equation}
\begin{array}{l}
\sigma(p\bar{p}\to J/\Psi X,|y(J/\Psi|\le0.6)
\times Br(J/\Psi\to\mu\mu)\\\\
=240\pm1(stat)^{+35}_{-28}(syst)\,\mbox{nb},
\end{array}
\end{equation}
Taking advantage of the large azimuth coverage of its
muon system, the \D0 measured the differential cross
section of the inclusive $J/\Psi$ as a function of the
rapidity using $4.74\,\mbox{pb}^{-1}$. The distribution
is shown in Fig.~\ref{fig:xsec_Jpsi_D0}.

\section{Bottom quark Production Cross Section}
The Run\,I central $b$-quark production cross section 
measured by CDF and \D0 has a minimum transverse momentum 
cut on the $b$-hadrons due to the trigger requirements.
Since those measurements only explore a small fraction 
($\sim10\,\%$) of the $b$-hadron $p_T$ spectrum, it is not 
clear from the data  whether the excess over the theory 
is due to an overall discrepancy of the $b\bar{b}$ 
production rate, or it is caused by a shift in the spectrum
toward higher $p_T$. An inclusive measurement of bottom quark 
production over all transverse momentum can certainly help 
resolve this ambiguity.

\begin{figure}
\scalebox{0.4}{\includegraphics{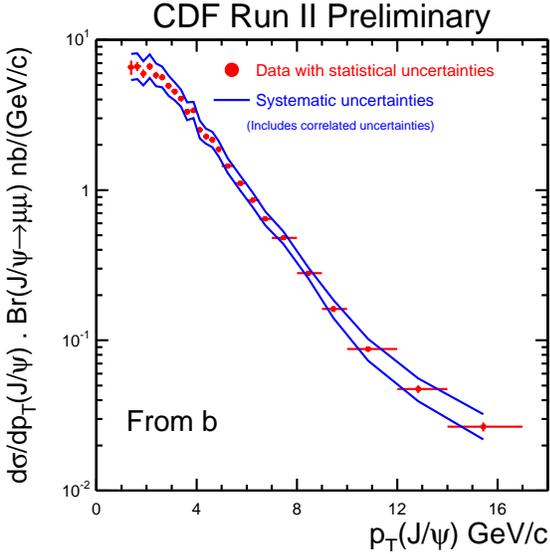}}
\caption{\label{fig:xsec_b_jpsi}{$b$-hadron differential 
cross section as a function of $p_T(J/\Psi)$.}}
\end{figure}

Notice that large fraction of $b$-hadron $H_b$ decays to 
$J/\Psi$ final states, where $H_b$ denotes both hadron and
anti-hadron.  
CDF performed an analysis to extract the inclusive 
$b$-hadron cross section from the measured inclusive $J/\Psi$ 
production cross section. Due to the long live lifetime of the 
$b$-hadrons, the vertex of the $J/\Psi$ decayed from a 
$b$-hadron (secondary $J/\Psi$) is usually hundreds microns 
away from the primary vertex, 
while for the prompt $J/\Psi$'s, which are directly produced or 
decayed from higher charmonium states, their vertex are at the 
proton-antiproton interaction point. As the result, we can 
statistically separate these two components by examining the 
projected $J/\Psi$ decay
distance along its transverse momentum. The measured $b$-fraction is 
then applied to the previously measured inclusive $J/\Psi$ cross
section to obtain the differential production cross section of 
$H_b\to J/\Psi X$ as a function of $p_T(J/\Psi)$. However, the 
above algorithm to extract $b$-fraction does not work for 
$J/\Psi$ candidate with low transverse momentum, because 
a $b$-hadron with low $p_T$ does not travel far away 
from the primary vertex in the transverse plane. Therefore a 
minimum $p_T(J/\Psi)\ge1.25\,\mbox{GeV}/c$ requirement is
imposed for the $b$-hadron cross section measurement as a function
of $p_T(J/\Psi)$ in order to have a reliable determination of 
$b$-fraction. The measured differential cross section result 
is shown in Fig.~\ref{fig:xsec_b_jpsi}. 

\begin{figure}
\scalebox{0.4}{\includegraphics{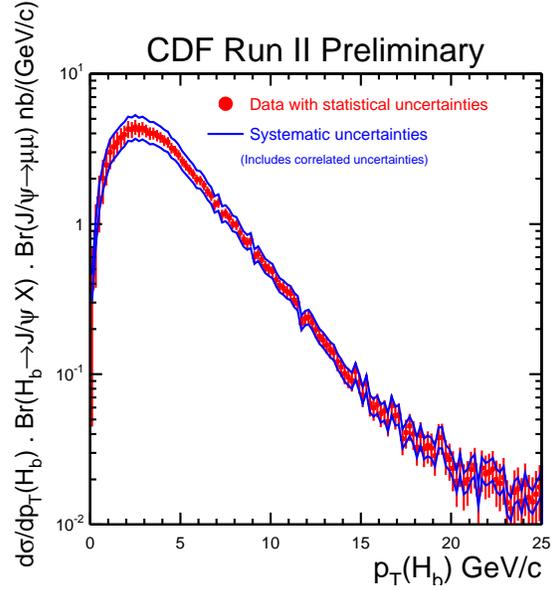}}
\caption{\label{fig:xsec_b}{$b$-hadron differential 
cross section as a function of $p_T(H_b)$ 
with $|y|\le 0.6$.}}
\end{figure}

Luckily, the $b$-hadron decayed at rest can still transfer 
a few ($\sim1.7$) $\mbox{GeV}/c$ transverse momentum to 
its $J/\Psi$ daughter because of the large $b$-hadron mass.
The knowledge of the secondary $J/\Psi$ production with transverse 
momenta less than $2.0\,\mbox{GeV}/c$ allows us to probe 
the $b$-hadrons with low transverse momenta down to zero.
Therefore using the measured $b$-hadron differential cross section 
with $1.25\le p_T(J/\Psi)\le 17.0\,\mbox{GeV}/c$, we 
are able to extract the
$b$-hadron differential cross section as a function of 
$p_T(H_b)$ down to $p_T(H_b)=0\,\mbox{GeV}/c$. To do so, we
perform a convolution that is based on the Monte Carlo 
template of the $b$-hadron transverse momentum distribution.
The unfolding procedure is then  repeated using the measured 
$b$-hadron production spectrum as the input spectrum for
the next iteration, until
the $\chi^2$ comparison between the input and output spectrum 
calculated after each iteration becomes stable. The measured
$b$-hadron cross section as a function of $p_T(H_b)$ is
plotted in Fig.~\ref{fig:xsec_b}.
The total $b$-hadron cross section is found to be 
\begin{equation}
\begin{array}{l}
\sigma(p\bar{p}\to H_b X,|y(H_b)|\le0.6)
\times Br(H_b\to J/\Psi X)\\\\
\times Br(J/\Psi\to\mu\mu)
=24.5\pm0.5(stat)^{+4.7}_{-3.9}(syst)\,\mbox{nb},
\end{array}
\end{equation}
The total single $b$-quark cross section is obtained by 
divided this measurement by two, the branching fractions, 
and the rapidity correction factors obtained from MC. 
We find:
\begin{equation}
\begin{array}{l}
\sigma(p\bar{p}\to \bar{b} X,|y(b)|\le0.6)\\\\
=18.0\pm0.4(stat)\pm3.8(syst)\,\mu\mbox{b},
\end{array}
\end{equation}
and
\begin{equation}
\begin{array}{l}
\sigma(p\bar{p}\to \bar{b} X,|y(b)|\le1)\\\\
=29.4\pm0.6(stat)\pm6.2(syst)\,\mu\mbox{b}.
\end{array}
\end{equation}

\section{$b$-jet Production Cross Section}
\begin{figure}
\scalebox{0.35}{\includegraphics{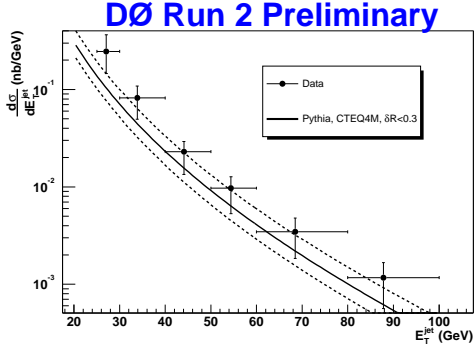}}
\caption{\label{fig:bjetcross}
Measured Run\,II $b$-jet cross section compared to 
theoretical prediction.}
\end{figure}

Using $3.4\,\mbox{pb}^{-1}$ of data, a
$b$-jet production cross section analysis has been performed
by \D0 . The candidate event is selected
by associating a muon track with a jet, which is defined with 
$R=\sqrt{\Delta\eta^2+\Delta\phi^2}$ cone algorithm.   
Notice the fact that the muon decayed from $b$ quark
has a higher transverse momentum with respect to 
the net momentum vector of combined muon and jet, 
the corresponding distribution is then used to fit the b-jet fraction.
The signal template is modeled from a $b\to\mu$ Monte
Carlo simulation, and the background template is extracted 
from 1.5 million QCD events. After correcting the muon and
jet reconstruction efficiency, and the calorimeter jet 
energy resolution, the $b$-jet cross section is obtained.
The preliminary result is shown in Fig.~\ref{fig:bjetcross}
compared to the theoretical prediction. This measurement is 
consistent with the previous Run\,I 
measurement~\cite{Abbott:2000iv}.

\section{Conclusion}
The understanding of the heavy flavor production is currently one
of the most important challenges faced by QCD. In this paper,
we present the most recent measurements from CDF and \D0 experiments.
Given large amount of data that have already been collected, rapid 
improvement of Tevatron performance, and further understanding of the
detector, we expect that much more precise measurements
on the heavy flavor production will be available in the near future.
At the same time, several other analyses have also been 
carried out to explore other aspects of heavy flavor production 
mechanism, such 
as the $b\bar{b}$ correlation and $c\bar{c}$ correlation 
during the production. 
In the next a few years, a combination of better experimental data
and improved theory should advance our knowledges of the 
heavy quarks production.


\begin{theacknowledgments}
We thank the CDF and \D0 collaboration for their helps
while preparing this paper. We also thank the conference
organizers for a wonderful meeting.

\end{theacknowledgments}


\bibliographystyle{aipproc}   


\IfFileExists{\jobname.bbl}{}
 {\typeout{}
  \typeout{******************************************}
  \typeout{** Please run "bibtex \jobname" to optain}
  \typeout{** the bibliography and then re-run LaTeX}
  \typeout{** twice to fix the references!}
  \typeout{******************************************}
  \typeout{}
 }

\end{document}